\documentclass[twocolumn,aps,showpacs,amsmath,amssymb,floatfix,superscriptaddress]{revtex4}
\usepackage{graphicx}
\usepackage{dcolumn}
\usepackage{bm}
\usepackage{hyperref}
\usepackage{latexsym}
\usepackage{float}
\usepackage{supertabular}
\usepackage{longtable}

\begin{document}
\title{On the meaning of the $h$-index} \author{S.~Redner}
\affiliation{Center for Polymer Studies and
and Department of Physics, Boston University, Boston, MA, 02215}

\begin{abstract}

  The $h$-index --- the value for which an individual has published at least
  $h$ papers with at least $h$ citations --- has become a popular metric to
  assess the citation impact of scientists.  As already noted in the original
  work of Hirsch and as evidenced from data of a representative sample of
  physicists, $\sqrt{c}$ scales as $h$, where $c$ is the total number
  citations to an individual.  Thus $\sqrt{c}$ appears to be equivalent to
  the $h$ index.  As a further check of this equivalence, the distribution of
  the ratio $s\equiv \sqrt{c}/2h$ for this sample is sharply peaked about 1.
  The outliers in this distribution reveal fundamentally different types of
  individual publication records.

\end{abstract}

\pacs{02.50.Ey, 05.40.-a, 05.50.+q, 89.65.-s}

\maketitle


What is the best way to assess the influence of scientific publications of
individual scientists?  Traditionally, this assessment has been based on the
number of publications of a scientist or the total number of citations
received.  However, in any creative endeavor, such as physics research, the
total amount of output is not necessarily the right metric for productivity.
In fact, L. D.  Landau himself\cite{LL} kept a list of physicists that were
ranked on a logarithmic scale of achievement.

Recently, Hirsch~\cite{H} introduced the $h$-index that attempts to capture
the overall impact of an individual's publication record researcher by a
single number.  The total number of publications can be misleading because an
individual could simply publish a large number of worthless articles.
Conversely, the total number of citations could also be misleading because an
individual might publish a single highly-cited article in a hot but transient
subfield but then nothing else of scientific value.  Such a citation record
may not be valuable as that of someone who steadily authors good publications
that are reasonably cited.

The idea underlying the $h$-index is that an equitable integral measure of
citation impact is provided by the value $h$, such that an individual has
published at least $h$ papers with at least $h$ citations.  It is obvious
that the $h$-index of a prolific author of trivial publications and that of a
researcher with a single great publication will be much less than someone who
publishes good papers at a steady rate.  Because of its obvious appeal, the
$h$-index has become a universally-used metric of overall citation impact.
As one example of the prominence of the $h$-index, it is immediately quoted
in Web of Science citation reports~\cite{ISI}.  Moreover, the original idea
of the $h$-index has spawned various of efforts to make the $h$-index more
``fair''~\cite{fair} by correcting for some of the obvious biases that are
part of the citation record, such as many co-authors, self-citations, role of
thesis advisor, {\it etc.}

However, as noted by Hirsch in his original publication~\cite{H}, the
$h$-index of an individual should scale as the square-root of the total
number citations to this individual.  This square-root scaling arises in the
most simple model of citations in which an individual publishes papers at a
constant rate and each publication is cited at a constant rate.  As a result,
the total number of citations grows quadratically with time while the
$h$-index grows linearly with time, {\it i.e.} $\sqrt{c}$ scales linearly
with $h$.  Here, we test this observation for a representative sample of 255
condensed-matter and statistical physics theorists in North America and
Europe.

\begin{figure}[ht]
 \vspace*{0.cm}
  \includegraphics*[width=0.45\textwidth]{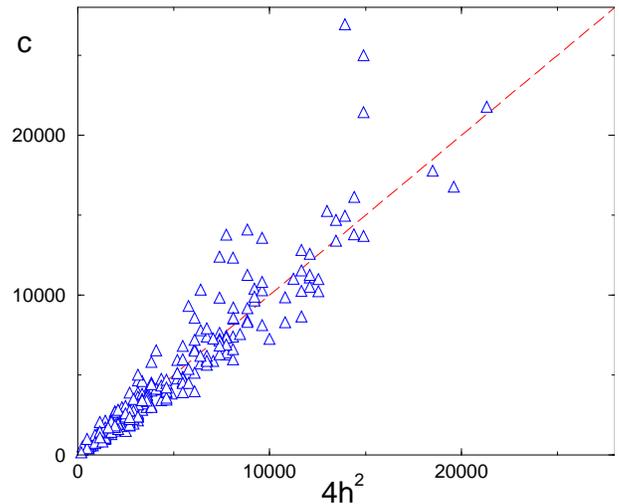}
  \caption{Plot of $c$ versus $4h^2$ for the 255 individuals in the dataset.
    The line $c=4h^2$ is shown dashed.}
\label{hsq-vs-c}
\end{figure}

The data was obtained by starting with the names of well-known
condensed-matter and statistical physics theorists and looking up their
citation record in the ISI Web of Science.  By scanning at the author lists
of the top-cited publications of these initial authors, the initial list of
authors was extended to their main collaborators, and then to collaborators
of collaborators, {\it etc.}  After about 250 people, it became difficult to
find new people or people who could be unambiguously resolved in the ISI
database with the limited knowledge of the author.  Primarily because of
limited personal knowledge, the dataset also under-represents junior people.
Moreover, because the Boston University institutional subscription for ISI
extends only to citations after 1973, individuals who began publishing before
this year were excluded to avoid the use of incomplete citation data for
their publications.  The data were gathered during a two-day period January
30--31, 2010 between updates of the science citation index database.

If $\sqrt{c}$ scales linearly with $h$, then a plot of these two quantities
should yield a straight line.  Fig.~\ref{hsq-vs-c} illustrates this behavior
for all the individuals in the dataset.  To highlight the outliers to the
linear behavior that will be discussed below, Fig.~\ref{hsq-vs-c} actually
shows $c$ versus $4h^2$.  A linear least-squares fit to all the data of
$\sqrt{c}$ versus $2h$ gives a best fit value of the slope $s\equiv
\sqrt{c}/2h$ of $s\approx 1.045$.  The data therefore suggest that $\sqrt{c}$
is essentially equivalent to the $h$-index, up to an overall factor that is
close to 2.

\begin{figure}[ht]
 \vspace*{0.cm}
  \includegraphics*[width=0.45\textwidth]{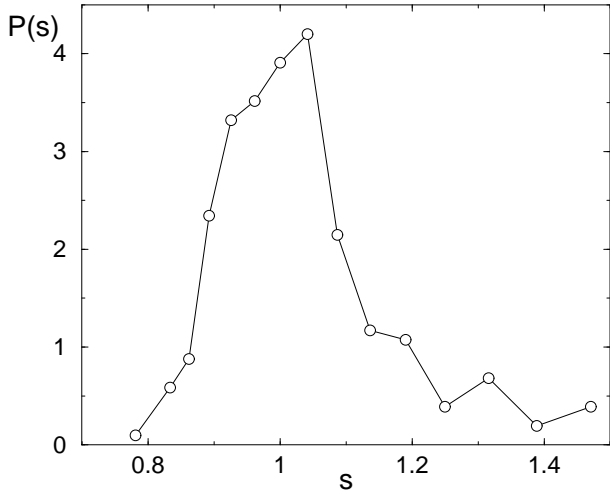}
  \caption{Plot of the probability density $P(s)$ that an individual is
    characterized by a value $s=\sqrt{c}/2h$.}
\label{dist}
\end{figure}

As a further test of the linearity of the dependence of $h$ versus
$\sqrt{c}$, the quantity $s=\sqrt{c}/2h\,$ is computed for each individual in
the dataset of 255 physicists and the resulting distribution, $P(s)$, is
shown in Fig.~\ref{dist}.  This distribution is fairly symmetric and most of
the data lies within the range $|s-1|<0.2$.  The tightness of the range of
$s$ again suggests that the relation $\sqrt{c}=2h$ accounts for most of the
citation data.

The outliers in the distribution $P(s)$ with $s<1$ and with $s>1$ are
particularly interesting.  In the scatter plot of $c$ versus $4h^2$ in
Fig.~\ref{hsq-vs-c}, consider first the outliers with $s<1$ --- data points
that lie below the diagonal.  As illustrated in table~\ref{S}, the citation
patterns of best-cited publications for the individuals with the smallest ten
values of $s$ are remarkably similar even though the $h$ indices of this
group of researches ranges over a factor of more than two.  In particular,
the difference in the number of citations of successive top-cited papers is
relatively small in all cases.  For example, the ratio of the number of
citations to the top-cited and third-cited paper for each individual is in
the range 1.025--2.072.

For the twenty individuals with the largest values of $s$, the citations
patterns are also quite similar within this subpopulation.  Almost all have
one (or a few) papers whose citations are a substantial factor larger than
their second-ranked paper.  For example, the largest ratio between the number
of citations of the top-cited and third-cited paper is now 10.03.  This wide
disparity arises because each individual in this subpopulation (co)-authored
one (or a few) famous publications whose citation frequency outstrips the
remaining publications.  Among the individuals that (co)-author these famous
publications, there are three clearly-defined situations: (i) individuals
that wrote a ground-breaking publication on their own or were the driver of
publication with a junior co-author, (ii) those that collaborated with a more
senior author in a famous publication, and (iii) those whose famous
publication was a particularly timely or authoritative review article.

\begin{widetext}

\begin{longtable}
{|p{0.3in}|p{0.35in}|p{0.4in}|p{0.1in}|p{0.3in}|p{0.3in}|p{0.3in}|p{0.3in}|p{0.3in}|p{0.3in}|p{0.3in}|p{0.3in}|p{0.3in}|p{0.3in}|}
\endhead
\caption{List of the top-10 cited publications of the individuals with the
  ten smallest values of $s=\sqrt{c}/2h$.  The first three columns give the
  $h$-index, the total number of citations $c$, and $s=\sqrt{c}/2h$.  The
  columns labeled $c_i$ for $i=1,2,\ldots, 10$ are the respective number of
  citations of the 10
  best-cited papers for each individual.}\label{S}\\
\hline $h$ & $c$ & $\sqrt{c}/2h$ && $c_1$& $c_2$& $c_3$& $c_4$& $c_5$& $c_6$& $c_7$& $c_8$& $c_9$& $c_{10}$\\ \hline 
25 & 1510 & 0.777 && 84 & 81 & 62 & 48 & 46 & 43 & 42 & 39 & 37 & 36\\ \hline
39 & 3983 & 0.809 && 260 & 177 & 144 & 127 & 126 & 92 & 91 & 90 & 89 & 85\\ \hline
18 & 853 & 0.811 && 172 & 153 & 83 & 72 & 49 & 39 & 36 & 35 &33  &23 \\ \hline
27 & 1966 & 0.821 && 197 & 191 & 139 & 110 & 66 & 66 & 52 & 51 & 48 & 44\\ \hline
26 & 1854 & 0.828 && 83 & 81 & 81 & 72 & 70 & 68 & 63 & 56 & 55 & 52\\ \hline
28 & 2169 & 0.832 && 100 & 95 & 92 & 89 & 83 & 75 & 73 & 67 & 64 & 64\\ \hline
19 & 1002 & 0.833 && 68 & 66 & 64 & 56 & 51 & 51 & 50 & 43 & 42 & 39\\ \hline
26 & 1879 & 0.833 && 148 & 141 & 84 & 76 & 75 & 65 & 64 & 62 & 56 & 54\\ \hline
23 & 1480 & 0.836 && 94 & 64 & 64 & 62 & 58 & 51 & 49 & 47 & 46 & 42\\ \hline
54 & 8209 & 0.839 && 316 & 297 & 285 & 199 & 198 &198  &181 &177& 162& 153 \\ \hline
\end{longtable}
\newpage
\begin{longtable}
{|p{0.3in}|p{0.35in}|p{0.4in}|p{0.1in}|p{0.3in}|p{0.3in}|p{0.3in}|p{0.3in}|p{0.3in}|p{0.3in}|p{0.3in}|p{0.3in}|p{0.3in}|p{0.3in}|}
\endhead
\caption{List of the citation record of the individuals with the twenty
  largest values of $s=\sqrt{c}/2h$; the data format is the same as
  Table~\ref{S}.  Italicized entries denote review articles.}\label{HR}\\
\hline $h$ & $c$ & $\sqrt{c}/2h$ &&  $c_1$& $c_2$& $c_3$& $c_4$& $c_5$& $c_6$& $c_7$& $c_8$& $c_9$& $c_{10}$\\ \hline 
8 & 544 & 1.458 && 141 & 135 & 50 & 34 & 31 & 17 & 13 &13  & 8 &8 \\ \hline
11 & 1011 & 1.445 && 329 & 220 & 105 & 75 & 73 & 37 & 28 & 24 &24  &17 \\ \hline
20 & 3163 & 1.406 && 480 & 303 & 276 & 264 & 257 & 212 & 198 & 191 & 165 & 157\\ \hline
59 & 26937 & 1.391 && 2259 & 1830 & 1310 & 1220 & 784 &777 &606 &355 &54  &312 \\ \hline
44 & 13789 & 1.334 && 1824 & 1469 & 1393 & 1042 & 570 &560  &504  &480&327 &316 \\ \hline
17 & 2058 & 1.334 && 550 & 255 & 197 & 194 & 123 & 97 & 81 & 73 & 70 & 70\\ \hline
27 & 4903 & 1.297 && 2004 & 371 &316 &243 &157 &133 &114 &100 & 98 &97 \\ \hline
61 & 25003& 1.296 &&4461&  \textit{3778} & 1444 & 1333 & 1176 & 1104 &1101 &835 &651 &400\\ \hline
43 & 12403 & 1.295 && 4148 & 1561 & 551 & 495 &452  &405  &399  &339 &217  &214 \\ \hline
40 & 10347 & 1.271 && 2118 & 2004 & 857 &433  &292  & 281 & 274 &238  &223  &221\\ \hline
38 & 9331 & 1.271 && 2721 & 828 & 530 &472 &466&451&324&271&205&178 \\ \hline
32 & 6537 & 1.263 && 1105 & 735 & 650 & 525 & 516 & 320 & 174& 154& 151& 138  \\ \hline
47 & 14090 & 1.263 &&  \textit{3232} & 815 & 699 & 620 &477&466&420&353&329&274\\ \hline
45 & 12347 & 1.235 && \textit{2357} & 765 & 641 & 563 &495 &462 &405 &377 &350  &322 \\ \hline
28 & 4660 & 1.219 && 2260 & 274 & 206 & 140 &116 &86&84&83&81&79 \\ \hline
19 & 2137 & 1.271 && 766 & 301 & 182 & 77 &74  &71  &61  &58  &43  &41 \\ \hline
61 & 21446 & 1.200 && 7014 & 1102 & 699 & 626 & 502 &427 &331&325&304&296 \\ \hline
15 & 1274 & 1.190 && 242 & 232 & 140 & 96 & 66 & 57& 48 & 41 & 34 & 33 \\ \hline
49 & 13582 & 1.189 && 3051 & 985 & 883 & 864 & 698 & 374 &349 &349&302  &241 \\ \hline
22 & 2732 & 1.188 && 569 & 343 & 271 & 192 & 165 & 98 & 96 & 90 & 72 & 63\\ \hline
39 & 8584 & 1.188 && 2260 & 980 & 658 & 451 &  296 & 289 & 269 & 149& 147&144 \\ \hline
22 & 2699 & 1.181 && 507 & 340 & 192 & 184 & 145 & 130 & 121 & 93 & 92 & 90\\ \hline
\end{longtable}
\end{widetext}

One basic conclusion from this study is that the square-root of the total
number of citations that an individual receives very nearly coincides with
twice his or her $h$-index.  A still an open question is why should
$\sqrt{c}$ provide the same integrated measure of the breadth and depth of an
individual's citation record as the $h$-index itself.  

A second conclusion is that it is possible to identify outstanding
researchers as the outliers above the diagonal in the scatter plot of
Fig.~\ref{hsq-vs-c}.  While there are roughly the same number of points below
the diagonal as above the diagonal, the above-diagonal points with roughly
9000 citations or greater are visually prominent and correspond to
individuals with seminal publications.  This simple characteristic appears to
provide a useful predictor of research excellence.

A final caveat: while the outliers discussed here correspond to researchers
with excellent publications to their credit, there are many examples of
excellent researchers that do not fit this outlier criterion.  It is
important to be aware of the limitations of using citations alone, or some
function of the number of citations, as a measure of research excellence.

\acknowledgments{I gratefully acknowledge financial support from the US
  National Science Foundation grant DMR0906504.  I also thank S. Dorogovtsev
  for initial correspondence that kindled my old interest in this subject and
  J. E. Hirsch for friendly correspondences and advice.}

\end{document}